\documentstyle[12pt]{article}
\input epsf

\begin{document}
\baselineskip=15pt
\newcommand{\x}{{\bf x}}
\newcommand{\y}{{\bf y}}
\newcommand{\z}{{\bf z}}
\newcommand{\bp}{{\bf p}}
\newcommand{\A}{{\bf A}}
\newcommand{\B}{{\bf B}}
\newcommand{\p}{\varphi}
\newcommand{\del}{\nabla}
\newcommand{\be}{\begin{equation}}
\newcommand{\ee}{\end{equation}}
\newcommand{\bq}{\begin{eqnarray}}
\newcommand{\eq}{\end{eqnarray}}
\newcommand{\ba}{\begin{eqnarray}}
\newcommand{\ea}{\end{eqnarray}}
\def\r{\nonumber\cr}
\def\hf{\textstyle{1\over2}}
\def\qr{\textstyle{1\over4}}
\def\Sc{Schr\"odinger\,}
\def\sc{Schr\"odinger\,}
\def\'{^\prime}
\def\>{\rangle}
\def\<{\langle}
\def\-{\rightarrow}
\def\dbd{\partial\over\partial}
\def\tr{{\rm tr}}
\def\hg{{\hat g}}
\def\ca{{\cal A}}
\def\pd{\partial}
\def\dl{\delta}

\begin{titlepage}
\vskip1in
\begin{center}
{\large Cancellation of divergences in ${\cal N}=4$ SYM/Type IIB
Supergravity correspondence.}
\end{center}
\vskip1in
\begin{center}
{\large David Nolland}

\vskip20pt

Department of Mathematical Sciences

University of Liverpool

Liverpool, L69 3BX, England

{\it nolland@liv.ac.uk}

\end{center}
\vskip1in
\begin{abstract}

\noindent Using Schr\"odinger functional methods, we show that in
the ${\cal N}=4$ SYM/Type IIB Supergravity correspondence the
renormalisation of the boundary Newton and gravitational constants
arising from bulk fields cancels when we sum over all the
Kaluza-Klein modes of Supergravity. This accords with the expected
finiteness of ${\cal N}=4$ SYM, and it is expected that other
renormalisations cancel in a similar way.

\end{abstract}

\end{titlepage}

The correspondence between ${\cal N}=4$ Super-Yang-Mills Theory
and Type IIB Supergravity/String Theory \cite{review} has been of
considerable importance in shedding light on both theories. Most
of the work that has been done to date has focussed on the
large-$N$ limit of the SYM theory, largely because the calculation
of string loops on the AdS background is not well understood.
However, many important subleading order effects in the large-$N$
expansion correspond to Supergravity loops and can be calculated.
An example of this is the Weyl anomaly, which receives
contributions at one loop from all of the Kaluza-Klein modes of
Supergravity \cite{us,us3}.

Loop effects in supergravity also renormalise the boundary
wave-functional, which according to the correspondence is
identified with the partition function of the boundary theory. But
the finiteness of the boundary theory leads us to expect that such
renormalisations should disappear when the full theory is taken
into account.

The purpose of this letter is to demonstrate this cancellation of
divergences for the contributions of bulk supergravity fields to
the renormalisation of the boundary Newton and cosmological
constants. When we sum over all the Kaluza-Klein modes of Type IIB
Supergravity compactified on $AdS_5\times S^5$ the contributions
to the renormalisation cancel, so that no renormalisation is
needed when all the bulk modes are taken into account.

The way in which this cancellation happens is significant. For
Ricci-flat boundaries the bulk metric is unaffected by introducing
boundary curvature and the cancellation happens within
supermultiplets. For non-Ricci flat boundaries the bulk metric
acquires an extra factor (though the bulk metric satisfies the
same Einstein equations) and the cancellation requires an analytic
regularisation of the infinite sum over Kaluza-Klein modes. The
latter case demonstrates that it is insufficient to consider only
the consistent truncation to the massless multiplet of Type IIB
Supergravity. So, for example, for a non-Ricci flat boundary, the
calculation of the anomaly \cite{us,us3} to the truncated spectrum
of \cite{warner} fails to produce the expected subleading
correction to the coefficient $c$ for the infra-red fixed point of
the RG flow driven by adding certain mass terms to the ${\cal
N}=4$ Super-Yang-Mills theory to break the supersymmetry down to
${\cal N}=1$. For a Ricci-flat boundary, however, the truncated
spectrum gives the correct result \cite{us4}.

We expect that other renormalisations due to bulk interactions
cancel in a similar way to the cancellations that we describe
here.

The bulk $AdS_5$ metric giving a general Einstein metric $\hat g$
on the boundary is

\begin{equation}
ds^2 = G_{\mu\nu}\,dX^\mu\,dX^\nu=dr^2 + z^{-2}¥\, e^{\rho}
\hg_{ij}(x)\, dx^i dx^j \, ,\quad e^{\rho/2}= 1-C\,z^{2}\,, \quad
C={l^2 {\hat R} \over 48}\quad z=\exp(r/l),\label{ads1}
\end{equation}

and a regularisation is introduced by putting the boundary at
$z=\tau=\exp(r_0/l)$. Consider a scalar field of mass $m$
propagating in this metric; it has the action

\bq S_\phi&=&{1\over 2}\int d^5 X{\sqrt
G}\left(G^{\mu\nu}\partial_\mu\phi\,
\partial_\nu\phi+m^2\phi^2\right)\nonumber \\
&=&{1\over 2}\int {d^4x\,dr\over
z^4}{\sqrt\hg}\,e^{2\rho}\,\left(\dot\phi^2 +z^2
e^{-\rho}\hg^{ij}\partial_i\phi\,\partial_j\phi+m^2\phi^2\right),\label{sca}
\eq

with the dot denoting differentiation with respect to $r$. The
norm on fluctuations of the field from which the functional
integral volume element can be constructed is

\be ||\delta\phi||^2=\int d^5 X{\sqrt G}\,\delta\phi^2 =\int
{d^4x\,dr\over z^4}{\sqrt\hg}\,e^{2\rho}\,\delta\phi^2\,, \ee

and it is convenient to redefine the field by setting
$\phi=z^2\,e^{-\rho}\varphi$ to make the `kinetic' term in the
action into the standard form. The action becomes

\be S_\phi={1\over 2}\int
{d^4x\,dr}{\sqrt\hg}\,\left(\dot\varphi^2+z^2 e^{-\rho}\varphi
\left(\Box+{{\hat R}\over 6}\right)\varphi+ \left(m^2+{4\over
l^2}\right)\varphi^2\right),\ee

where we discarded a boundary term that is eventually sent to zero
by wave-function renormalisation \cite{bcs}. According to the
AdS/CFT correspondence the boundary partition function is

\be \Psi=\int {\cal
D}\varphi\,e^{-S_{\varphi}}\Big|_{\varphi(r=r_0)=\hat\varphi}
\equiv e^{W[\hat\varphi,\hat g]} \qquad W[\hat\varphi]=F+{1\over
2}\int d^4 x\,\sqrt{\hg}\,\hat\varphi\Gamma\,\hat\varphi,\ee where
$\Gamma$ is a differential operator and $F$ is the free energy of
the scalar field. This satisfies a functional Schr\"odinger
equation that can be read off from the action

\be {\pd \over \pd r_0}\,\Psi=-{1\over 2} \int d^4x \sqrt{\hg}\,
\Big\{-\hg^{-1}\, {\dl^2 \over \dl\varphi^2} + \tau^2
e^{-\rho}\varphi \left(\Box+{{\hat R}\over
6}\right)\varphi+\left(m^2+{4\over l^2}\right) \varphi^2
\Big\}\,\Psi\,. \label{sch2} \ee which implies that \be {\pd \over
\pd r_0}\Gamma=\Gamma^2-\tau^2 e^{-\rho}\left(\Box+{{\hat R}\over
6}\right)-\left(m^2+{4\over l^2}\right),\qquad {\pd \over \pd
r_0}F={1\over 2}{\rm Tr}\, \Gamma.\label{gam} \ee

We can solve for $\Gamma$ in powers of the differential operator
by expanding \be \Gamma=\sum_{n=0}^\infty b_n(r_0) \left(\Box+\hat
R/6\right)^n\,,\label{ex} \ee so that \be b_0=-\sqrt{m^2+{4\over
l^2}}, \ee (we take the minus sign to give a normalisable
wave-functional). The other coefficients in (\ref{ex}) vanish as
the cut-off, $r_0$ is taken to $-\infty$.

The free energy can be regulated with a Seeley-de Witt expansion
of the heat-kernel

\be {\rm Tr}\, \Gamma=\sum_{n=0}^\infty b_n(r_0)\,\left(-{\pd
\over \pd s} \right)^n {\rm Tr}\,\exp \left(-s\left(\Box+\hat
R/6\right)\right) \ee  \be {\rm Tr}\,\exp \left(-s\left(\Box+\hat
R/6\right)\right) =\int d^4 x\,{\sqrt\hg}{1\over
16\pi^2s^2}\left(a_0+s\,a_1(x)
+s^2\,a_2(x)+s^3\,a_3(x)+...\right)\label{sdw} \ee

where to remove the regulator we take the proper-time separation
$s$ to zero, and $r_0\to-\infty$. The only surviving contributions
come from $a_0$, $a_1$ and $a_2$. The $a_2$ contribution is finite
and determines the Weyl anomaly \cite{us2}, but the $a_0$ and
$a_1$ contributions diverge and renormalise the boundary
cosmological and Newton's constants respectively.

Now we derived the Schr\"odinger equation for a scalar field, but
with a little work, we can find a similar equation for all the
fields of Supergravity, with the coefficients appearing in
(\ref{sdw}) being the appropriate coefficients for a conformal
field of the appropriate spin. Details of this will be given in
\cite{us2}. It makes sense to consider the same proper-time
separation for all the fields (this is inevitable if we rewrite
(\ref{sdw}) in a superfield formalism) and so it makes sense to
sum over the $a_0$ and $a_1$ coefficients of all the fields in the
bulk spectrum, in order to determine the overall renormalisation.

The divergent coefficients of $a_0$ and $a_1$ are proportional to
$\sqrt{l^2m^2+4}=\Delta-2$. In Table 1 we list the values of
$\Delta-2$ for the fields in the bulk spectrum, originally worked
out in \cite{Kim}. The multiplets are labelled by an integer
$p\ge2$ and live in representations of $SU(4)$. The $a_0$
coefficients are given by $(-1)^{2\sigma}{\rm Tr} 1$ where
$\sigma$ is the spin of the field. As a result, cancellation
within supermultiplets is guaranteed by the presence of equal
numbers of bosonic and fermionic modes.

The $a_1$ coefficients are given by ${1\over6}R{\rm Tr} 1-{\rm Tr
}E$, where $-\del^2-E$ is the operator associated with the
conformally coupled six-dimensional field (this is the operator
appearing in the heat-kernel). For a conformally coupled scalar,
fermion and gauge-fixed vector field this gives $a_1=0,$ $\hat
R/3$, and $-2\hat R/3$ respectively.

If we sum the coefficient of $a_1$ over all the fields of the
theory, the contribution to (\ref{sdw}) can be written as

\be \int d^4 x\,{\sqrt\hg}{1\over 16\pi^2s}\sum_p(\Delta-2)
a_1(x), \ee and denoting the values of $a_{1}$ for the fields
$\phi$, $\psi$, $A_\mu$, $A_{\mu\nu}$, $\psi_\mu$, $h_{\mu\nu}$ by
$ s,f,v,a,r, $ and $g$ respectively we have \bq \left(\sum
(\Delta-2)a_1\right)_{p\ge 4}&=&
(-4s+4a+r+f+2v){p\over 3}+\nonumber\\
(-105s-g-26a-8r-72f-48v){p^{3}\over 12}
&+&(16v+20f+10a+4r+25s+g){p^{5}\over 12} \eq whilst for the $p=3$
multiplet we have \be \left(\sum (\Delta-2)a_1\right)_{p=3}=
244f+18g+266s+218v+148a+64r\,. \ee The $p=2$ multiplet contains
gauge fields requiring the introduction of Faddeev-Popov ghosts,
whose parameters are listed in Table 2. This gives \be \left(\sum
(\Delta-2)a_1\right)_{p=2}=12 v- 6 s +6 r+6f +2g +12a \ee We have
to deal with the sum over multiplets labelled by $p$. We will
evaluate this divergent sum by weighting the contribution of each
supermultiplet by $z^p$. The sum can be performed for $|z|<1$, and
we take the result to be a regularisation of the weighted sum for
all values of $z$. Multiplying this by $1/(z-1)$ and integrating
around the pole at $z=1$ gives a regularisation of the original
divergent sum. This yields \be \sum (\Delta-2)a_1=8s+4f+2v
\label{tot}\ee which remarkably depends only on the heat-kernel
coefficients of fields in the Super-Yang-Mills theory. By
decomposing a five-dimensional vector into longitudinal and
transverse pieces and solving the Schr\"odinger equation for them,
it can be seen that the heat-kernel coefficient for a vector
field, $v$, is related to that for a four-dimensional gauge-fixed
Maxwell field, $v_0$, as $v = v_0 + 2s - 2s_0$ where $s_0$ is the
coefficient for a minimally coupled four-dimensional scalar
(Faddeev-Popov ghost), showing $v-2s = v_0 -2s_0 = g_v$
\cite{us2}. If we substitute the values for the $a_1$ coefficients
we see that the sum (\ref{tot}) vanishes, so that there is no
overall renormalisation of the boundary Newton's constant.

As emphasised earlier other renormalisations arising from
interactions of bulk Supergravity fields should follow a similar
pattern. If the boundary is taken to be Ricci-flat, we expect to
observe a cancellation within supermultiplets, whereas if the
boundary is non-Ricci flat we would need an additional
regularisation of the sum over Kaluza-Klein modes such as we have
made use of here.

\begin{table}[b]
\begin{center}
\caption{Mass spectrum. The supermultiplets (irreps of U(2,2/4))
are labelled
   by the integer $p$. Note that the doubleton ($p=1$) does not appear in
the
   spectrum. The $(a,b,c)$ representation of $SU(4)$ has dimension
   $(a+1)(b+1)(c+1)(a+b+2)(b+c+2)(a+b+c+3)/12$, and a subscript $c$
indicates
   that the representation is complex. (Spinors are four component Dirac
   spinors in $AdS_5$).}
\label{spec} \vskip .3cm
  \begin{tabular}{|cccc|}
\hline

     Field  & $SO(4)$ rep$^{\rm n}$ & $SU(4)$ rep$^{\rm n}$ &
$\Delta-2$      \\

  \hline

$\phi^{(1)}$ & $(0,0)$ & $(0,p,0)$ & $p-2$,\quad $p\ge2$ \\
$\psi^{(1)}$ & $(\hf,0)$ & $(0,p-1,1)_c$ & $p-3/2$,\quad $p\ge2$ \\
$A_{\mu\nu}^{(1)}$ & $(1,0)$ & $(0,p-1,0)_c$ & $p-1$,\quad
$p\ge2$ \\
\hline $\phi^{(2)}$ & $(0,0)$ & $(0,p-2,2)_c$ & $p-1$,\quad
$p\ge2$
\\
$\phi^{(3)}$ & $(0,0)$ & $(0,p-2,0)_c$ & $p$,\quad $p\ge2$
\\
$\psi^{(2)}$ & $(\hf,0)$ & $(0,p-2,1)_c$ & $p-1/2$,\quad $p\ge2$ \\
$A_\mu^{(1)}$ & $(\hf,\hf)$ & $(1,p-2,1)$ & $p-1$,\quad $p\ge2$
\\
$\psi_\mu^{(1)}$ & $(1,\hf)$ & $(1,p-2,0)_c$ & $p-1/2$,\quad
$p\ge2$
\\
$h_{\mu\nu}$ & $(1,1)$ & $(0,p-2,0)$ & $p$,\quad $p\ge2$ \\

\hline
$\psi^{(3)}$ & $(\hf,0)$ & $(2,p-3,1)_c$ & $p-1/2$,\quad $p\ge3$ \\
$\psi^{(4)}$ & $(\hf,0)$ & $(0,p-3,1)_c$ & $p+1/2$,\quad $p\ge3$ \\
$A_\mu^{(2)}$ & $(\hf,\hf)$ & $(1,p-3,1)_c$ & $p$,\quad
$p\ge3$ \\
$A_{\mu\nu}^{(2)}$ & $(1,0)$ & $(2,p-3,0)_c$ & $p$,\quad $p\ge3$
\\
$A_{\mu\nu}^{(3)}$ & $(1,0)$ & $(0,p-3,0)_c$ & $p+1$,\quad
$p\ge3$ \\
$\psi_\mu^{(2)}$ & $(1,\hf)$ & $(1,p-3,0)_c$ & $p+1/2$,\quad
$p\ge3$
\\
\hline
$\phi^{(4)}$ & $(0,0)$ & $(2,p-4,2)$ & $p$,\quad $p\ge4$ \\

$\phi^{(5)}$ & $(0,0)$ & $(0,p-4,2)_c$ & $p+1$,\quad $p\ge4$
\\
$\phi^{(6)}$ & $(0,0)$ & $(0,p-4,0)$ & $p+2$,\quad $p\ge4$ \\
$\psi^{(5)}$ & $(\hf,0)$ & $(2,p-4,1)_c$ & $p+1/2$,\quad $p\ge4$ \\
$\psi^{(6)}$ & $(\hf,0)$ & $(0,p-4,1)_c$ & $p+3/2$,\quad $p\ge4$ \\
$A_\mu^{(3)}$ & $(\hf,\hf)$ & $(1,p-4,1)$ & $p+1$,\quad $p\ge4$
\\
\hline
  \end{tabular}
  \end{center}
\end{table}

\begin{table}[t]
\begin{center}
\caption{Decomposition of gauge fields for the massless
multiplet.} \label{ghosts} \vskip .3cm
\begin{tabular}{|c|cccc|}
\hline
Original field & Gauge fixed fields & $\Delta-2$ & $R_{{ij}}=0$:& Constant $R$:\\
    &  &  & $180 a_{2}/R_{ijkl}R^{ijkl}$ & $180a_2/R^2$\\
\hline
$A_\mu$ & $A_i$ & 1 & -11 & 29/3\\
({\bf 15} of $SU(4)$)       & $A_0$ & 2 & 1 & -1/12\\
         & $b_{FP}$, $c_{FP}$ & 2 & -1 & 1/12\\
\hline
$\psi_\mu$ & $\psi_i^{\rm irr}$ & 3/2 & -219/2 & -61/4\\
&            $\gamma^i\psi_i$ & 5/2 & 7/2 & -11/12\\
({\bf 4} of $SU(4)$)  & $\psi_0$ & 5/2 & 7/2 & -11/12\\
& $\lambda_{FP}$, $\rho_{FP}$ & 5/2 & -7/2 & 11/12\\
& $\sigma_{GF}$ & 5/2 & -7/2 & 11/12\\
\hline
$h_{\mu\nu}$ & $h_{ij}^{\rm irr}$ & 2 & 189 & 727/4\\
($SU(4)$ singlet) & $h_{0i}$ & 3 & -11 & 29/3\\
& $h_{00}$, $h_\mu^\mu$ & $\sqrt{12}$& 1 & -1/12\\
& $B^{FP}_0$,$C^{FP}_0$ &$\sqrt{12}$& -1 & 1/12\\
& $B^{FP}_i$,$C^{FP}_i$ & 3 & 11 & -29/3\\
\hline
   \end{tabular}
  \end{center}
\end{table}

\end{document}